\documentclass[12pt,indentfirst]{article}

\usepackage{jsr}

\def\cl#1{\begin{center}\Large #1\end{center}}
\def\ce#1{\centerline{#1}}

\def\sh{\hat{\sigma}}
\def\shs{\sh^2}

\begin{document}

\baselineskip=18pt

\cl{\bf Simple Confidence Intervals for MCMC Without CLTs}

\ce{\jeff, October 24, 2016 \ (last revised January 9, 2017)}

\bigskip
\newpartitle{Summary}  This short note argues that 95\% confidence intervals
for MCMC estimates can be obtained even without establishing a CLT,
by multiplying their widths by 2.3.

\section{Introduction}

Markov chain Monte Carlo
(MCMC) algorithms are very widely used to estimate of expected values
in a variety of settings, especially for Bayesian inference (see e.g.\
Brooks et al., 2011, and the many references therein).

It has been pointed out by various authors (e.g.\ Jones and Hobert,
2001; Flegal et al., 2008) that in addition to providing an estimate,
it is also important to quantify the {\it error} in the estimate,
hopefully by providing {\it confidence intervals} for the value
being estimated.

Such error estimation and confidence intervals are
usually obtained via Markov chain Central Limit Theorems
(CLTs), see e.g.\ Tierney (1994, Theorem~4), Chan and Geyer (1994),
Jones (2004), Roberts and Rosenthal (2004),
and Jones et al.\ (2006).  Indeed, CLTs are often
considered {\it essential} for this purpose, e.g.\
Jones (2007, p.~131) writes ``The CLT is the basis of all
error estimation in Monte Carlo''.  However, establishing CLTs
for MCMC requires the verification of challenging properties like
geometric ergodicity, which is often difficult in applied problems.
This makes confidence intervals harder to obtain in MCMC applications.

In this short note, we show (Theorem~\ref{mainthm})
that for typical MCMC applications,
as long as the asymptotic variance
can be estimated, a confidence interval (or at least an {\it
upper-bound} on a confidence interval) can be obtained quite {\it
simply}, via Chebychev's inequality, without requiring any sort of
CLT or distributional convergence at all.

\section{Assumptions}

Let $\{X_n\}$ be a Markov chain on a state space $\X$ which converges
to a target distribution~$\pi$.
Let $h:\X\to\IR$ be some functional, and assume we wish to estimate
the stationary expected value of $h$, i.e.\
$\pi(h) := \int h(x) \, \pi(dx)$, by the usual MCMC estimate,
$e_n={1 \over n} \sum_{i=1}^n h(X_i)$.

In typical MCMC applications,
the estimate $e_n$ will have variance $O(1/n)$ and bias $O(1/n)$
(see e.g.\ page 21 of Geyer, 2011).
Consistent with this, we assume:

\medskip \noindent \bf (A1) \rm
{\it (Order $1/n$ variance.)}
The limit $V := \lim_{n\to\infty} n \Var(e_n)$ exists and is in $(0,\infty)$.
\medskip

\medskip \noindent \bf (A2) \rm
{\it (Smaller-order bias.)}
$\lim_{n\to\infty} n^{1/2} |\E(e_n) - \pi(h)| \, = \, 0$.
\bigskip

We also require an estimator of the asymptotic variance value
$V$.  Such estimators are quite common, and can be obtained in many
different ways, including repeated runs, integrated autocorrelation
times, batch means, window estimators, regenerations, and more; see
e.g.\ Section~3 of Geyer (1992), Hobert et al.\ (2002), Jones et al.\
(2006), H\"aggstr\"om and Rosenthal (2007), etc.  We thus assume:

\medskip \noindent \bf (A3) \rm
{\it (Variance estimator.)}
There is an estimator $\shs_n$
with $\lim_{n\to\infty} \shs_n = V$ in probability.

\section{Main Result}

Under the above mild assumptions, our result is as follows:

\pnew{Theorem}
\plabel{mainthm}
Assume (A1)--(A3) above, fix $0<\alpha<1$ and $\epsilon>0$,
and define the interval
$$
I_{n,\epsilon} \ := \
\Big( e_n - n^{-1/2} \sh_n \alpha^{-1/2} (1+\epsilon),
	\ e_n + n^{-1/2} \sh_n \alpha^{-1/2} (1+\epsilon) \Big)
\, .
$$
Then
$$
\liminf_{n\to\infty} \P\Big(\pi(h) \in I_{n,\epsilon}\Big)
		\ \ge \ 1 - \alpha
\, ,
$$
i.e.\ the interval $I_{n,\epsilon}$
includes the true expected value
$\pi(h)$ with asymptotic probability at least $1-\alpha$,
i.e.\ $I_{n,\epsilon}$ has asymptotic coverage probability
at least $1-\alpha$.

\medskip Theorem~\ref{mainthm} may be interpreted as saying that
the interval $I_{n,\epsilon}$ {\it contains} an asymptotic
$(1-\alpha)$-confidence interval for $\pi(h)$,
i.e.\ it is an overly-conservative confidence interval.
Since the main purpose of MCMC confidence intervals is to provide
approximate {\it guarantees} for estimates, this conservativeness
is not a major limitation.

Most commonly, the significance level $\alpha=0.05$.
In that case, the usual CLT-derived 95\% asymptotic confidence interval for
$\pi(h)$ would be given by
$[e_n - 1.96 \, \sh_n / \sqrt{n}, \ e_n + 1.96 \, \sh_n / \sqrt{n}]$.
By contrast, taking $\alpha=0.05$ and $\epsilon=0.001$,
our interval is computed to be
$I_{n,\epsilon}
= [e_n - 4.48 \, \sh_n / \sqrt{n}, \ e_n + 4.48 \, \sh_n / \sqrt{n}]$.
So, Theorem~\ref{mainthm} can be interpreted as saying that even without
establishing a Markov chain CLT,
the usual MCMC asymptotic
95\% confidence interval still applies, except with
``1.96'' replaced by ``4.48'', i.e.\ multiplying by just under 2.3
(and with the asymptotic coverage probability being
$\ge 95\%$ instead of exactly 95\%, i.e.\ being overly conservative).
Given the difficulty of establishing CLTs for MCMC algorithms, it
seems easier to instead simply multiply the confidence interval
width by 2.3.

\section{Proof of Theorem~\ref{mainthm}}

For any $a_n>0$, we have
by the triangle inequality that
$$
\P\Big(|e_n-\pi(h)| \ge a_n\Big)
\ = \ \P\Big(\Big|\Big(e_n-\E(e_n)\Big)+\Big(\E(e_n)-\pi(h)\Big)\Big|
	\ge a_n\Big)
$$
$$
\ \le \ \P\Big(|e_n-\E(e_n)| + |\E(e_n)-\pi(h)| \ge a_n\Big)
$$
$$
\ = \ \P\Big(|e_n-\E(e_n)| \ge a_n - |\E(e_n)-\pi(h)|\Big)
\, .
$$
Hence, if
$$
a_n - |\E(e_n)-\pi(h)| \ > \ 0
\, ,
\eqno(*)
$$
then by Chebychev's inequality (e.g.\ Rosenthal, 2006, Proposition~5.1.2),
$$
\P\Big(|e_n-\pi(h)| \ge a_n\Big)
\ \le \ \Var(e_n) \Big/ \Big(a_n - |\E(e_n)-\pi(h)|\Big)^2
\, .
$$

We now set
$a_n = \sqrt{ V/n\alpha }$.
Then by (A2), $\lim_{n\to\infty} |\E(e_n)-\pi(h)| \, / \, a_n = 0$.
Hence, $(*)$ is satisfied for all sufficiently large $n$, and
as $n\to\infty$, we have from the above and (A1) that
$$
\limsup_{n\to\infty} \P(|e_n-\pi(h)| \ge a_n)
\ \le \ \limsup_{n\to\infty} \ (V / n \, a_n^2)
\ = \ \limsup_{n\to\infty} \ (V / n \, (V/n\alpha))
\ = \ \alpha
\, .
$$

It remains to replace the true variance coefficient
$V$ by its estimator $\shs_n$.  For this, let $\epsilon>0$.  Then by (A3),
$\limsup_{n\to\infty} \P(\shs_n(1+\epsilon)^2 \le V)=0$.  Therefore,
\checkpage{3cm}
$$
\limsup_{n\to\infty} \P\Big( |e_n-\pi(h)|
		\ge n^{-1/2} \sh_n \alpha^{-1/2} (1+\epsilon) \Big)
\qquad\qquad\qquad\qquad\qquad\qquad
$$
$$
\ = \ \limsup_{n\to\infty} \P\Big(|e_n-\pi(h)|
		\ge \sqrt{\shs_n(1+\epsilon)^2/n\alpha}\Big)
\qquad\qquad\qquad\qquad
$$
$$
\ \le \ 
\limsup_{n\to\infty} \Big[
\P\left(|e_n-\pi(h)| \ge \sqrt{V/n\alpha} \quad {\rm or} \quad
\shs_n(1+\epsilon)^2 \le V\right) \Big]
$$
$$
\ \le \ 
\limsup_{n\to\infty} \Big[
\P\Big(|e_n-\pi(h)| \ge \sqrt{V/n\alpha}\Big)
+ \P\Big(\shs_n(1+\epsilon)^2 \le V\Big) \Big]
$$
$$
\ \le \ \alpha + 0
\ = \ \alpha
\, .
$$

Taking complements, we obtain that
$$
\liminf_{n\to\infty} \P\Big(|e_n-\pi(h)| < n^{-1/2} \sh_n \alpha^{-1/2}
(1+\epsilon) \Big)
\ \ge \ 1 - \alpha
\, .
$$
Finally, note that $|e_n-\pi(h)| < n^{-1/2} \sh_n \alpha^{-1/2} (1+\epsilon)$
if and only if $\pi(h) \in I_{n,\epsilon}$.
Hence, this completes the proof of Theorem~\ref{mainthm}.
\qed

\remark
The recent paper Atchad\'e (2016) also obtains
confidence intervals for MCMC without requiring CLTs.  However, its
results apply only to reversible chains, and require knowledge of the
spectrum of a complicated kernel~$\phi$, and proceed by establishing
convergence in distribution to a complicated generalised T-distribution
which appears to be difficult and challenging to work with, so they
cannot be described as ``simple''.

\ack I thank Jim Hobert for very useful comments, and thank the anonymous
referees for helpful reports.

\baselineskip=17pt

\section*{References}
\frenchspacing

Y.F.~Atchad\'e (2016), Markov Chain Monte Carlo
confidence intervals.  Bernoulli {\bf 22(3)}, 1808--1838.

S.~Brooks, A.~Gelman, G.L.~Jones, and X.-L.~Meng, eds.\ (2011),
Handbook of Markov chain Monte Carlo.
Chapman \& Hall / CRC Press.

K.S. Chan and C.J. Geyer (1994), Discussion of Tierney (1994).
Ann. Stat. {\bf 22}, 1747--1758.

J.M.~Flegal, M.~Haran, and G.L.~Jones (2008),
Markov Chain Monte Carlo: Can We Trust the Third Significant Figure?
Stat.\ Sci.\ {\bf 23(2)}, 250--260.

C.J.~Geyer (1992), Practical Markov chain Monte Carlo.  Stat. Sci.
{\bf 7}, 473-483.

C.J.~Geyer (2011), Introduction to Markov chain Monte Carlo.
Chapter~1 of Brooks et al.\ (2011).

O.~H\"aggstr\"om and J.S.~Rosenthal (2007),
On Variance Conditions for Markov Chain CLTs.
Elec. Comm. Prob. {\bf 12}, 454--464.

J.P. Hobert, G.L. Jones, B. Presnell, and J.S. Rosenthal (2002), On the
Applicability of Regenerative Simulation in Markov Chain Monte Carlo.
Biometrika {\bf 89}, 731--743.

G.L.~Jones (2004),
On the Markov chain central limit theorem.
Prob.\ Surv.\ {\bf 1}, 299--320.

G.L.~Jones, M.~Haran, B.S.~Caffo, and R.~Neath (2006), Fixed Width
Output Analysis for Markov chain Monte Carlo.
J. Amer. Stat. Assoc. {\bf 101}, 1537--1547.

G.L.~Jones (2007),
Course notes for STAT 8701: Computational Statistical Methods.

G.L. Jones and J.P. Hobert (2001), Honest exploration of intractable
probability distributions via Markov chain Monte Carlo. Statistical
Science {\bf 16}, 312--334.

G.O.~Roberts and J.S.~Rosenthal (2004),
General state space Markov chains and MCMC algorithms.
Prob.\ Surv.\ {\bf 1}, 20--71.

J.S.~Rosenthal (2006), A first look at rigorous probability theory,
2nd ed.  World Scientific Publishing Company, Singapore.

L. Tierney (1994), Markov chains for exploring posterior distributions
(with discussion).  Ann. Stat. {\bf 22}, 1701--1762.

\end{document}